\documentclass[a4paper]{article}

\usepackage{INTERSPEECH2021}
\usepackage{subcaption}
\usepackage{multirow}
\usepackage{array}
\usepackage{graphics}
\usepackage{todonotes}
\usepackage{graphicx}
\usepackage{hyperref}
\usepackage{cleveref}
\usepackage{booktabs}

\title{Does Audio Deepfake Detection Generalize?}
\name{Nicolas M. Müller$^{1}$, 
Pavel Czempin$^{2}$, 
Franziska Dieckmann$^{2}$,\\
Adam Froghyar$^{3}$,
Konstantin Böttinger$^{1}$
}
\address{
  $^1$Fraunhofer AISEC $\;\;$
  $^2$Technical University Munich $\;\;$
  $^3$why do birds GmbH
  }
\email{nicolas.mueller@aisec.fraunhofer.de}

\def\crnn/{\texttt{CRNNSpoof}}
\def\drn/{\texttt{DeepResidualNN}}
\def\meso/{\texttt{MesoNet}}
\def\mi/{\texttt{MesoInception}}
\def\lcnn/{\texttt{LCNN}}
\def\lcnna/{\texttt{LCNN-Attention}}
\def\lcnnl/{\texttt{LCNN-LSTM}}
\def\rn/{\texttt{RawNet2}}
\def\lstm/{\texttt{LSTM}}
\def\transformer/{\texttt{Transformer}}
\def\resnet/{\texttt{ResNet18}}
\def\rawgatst/{\texttt{RawGAT-ST}}
\def\RawPC/{\texttt{RawPC}}
\def\runs/{three}
\def\nummodels/{twelve}

\begin{document}

\maketitle

\begin{abstract}
Current text-to-speech algorithms produce realistic fakes of human voices, making deepfake detection a much-needed area of research.
While researchers have presented various deep learning models for audio spoofs detection, it is often unclear exactly why these architectures are successful: 
Preprocessing steps, hyperparameter settings, and the degree of fine-tuning are not consistent across related work. Which factors contribute to success, and which are accidental? 

In this work, we address this problem: 
We systematize audio spoofing detection by re-implementing and uniformly evaluating \nummodels/ architectures from related work. 
We identify overarching features for successful audio deepfake detection,
such as using \emph{cqtspec} or \emph{logspec} features instead of \emph{melspec} features, which improves performance by 37\% EER on average, all other factors constant.

Additionally, we evaluate generalization capabilities:
We collect and publish a new dataset consisting of $37.9$ hours of found audio recordings of celebrities and politicians, of which $17.2$ hours are deepfakes.
We find that related work performs poorly on such real-world data
(performance degradation of up to one thousand percent).
This could suggest that the community has tailored its solutions too closely to the prevailing ASVspoof benchmark and that deepfakes are much harder to detect outside the lab than previously thought.

\end{abstract}

\maketitle
\section{Introduction}

Modern text-to-speech synthesis (TTS) is capable of realistic fakes of human voices, also known as audio deepfakes or spoofs.
While there are many ethical applications of this technology, there is also a serious risk of malicious use.
For example, TTS technology enables the cloning of politicians' voices~\cite{AudioDee92:online,Zelensky}, which poses a variety of risks to society, including the spread of misinformation.

Reliable detection of speech spoofing can help mitigate such risks and is therefore an active area of research. 
However, since the technology to create audio deepfakes has only been available for a few years (see Wavenet \cite{oord2016wavenet} and Tacotron~\cite{tacotron1}, published in 2016/17),
audio spoof detection is still in its infancy.
While many approaches have been proposed (cf. ~\Cref{sec:related_work}),
it is still difficult to understand why some of the models work well: 
Each work uses different feature extraction techniques, preprocessing steps, hyperparameter settings, and fine-tuning. 
Which are the main factors and drivers for models to perform well? What can be learned in principle for the development of such systems?

Furthermore, the evaluation of spoof detection models has so far been performed exclusively on the ASVspoof dataset~\cite{todisco2019asvspoof, nautsch2021ASVspoof}, which means that the reported performance of these models is based on a limited set of TTS synthesis algorithms.
ASVspoof is based on the VCTK dataset~\cite{yamagishi2019vctk}, which exclusively features professional speakers and has been recorded in a studio environment, using a semi-anechoic chamber.
What can we expect from audio spoof detection trained on this dataset? Is it capable of detecting realistic, unseen, `in-the-wild' audio spoofs like those encountered on social media?

To answer these questions, this paper presents the following contributions:

\begin{itemize}
    \itemsep0em
    \item We reimplement \nummodels/ of the most popular architectures from related work and evaluate them according to a common standard. 
    We systematically exchange components to attribute performance reported in related work to either model architecture, feature extraction, or data preprocessing techniques. In this way, we identify fundamental properties for well-performing audio deepfake detection.
    \item To investigate the applicability of related work in the real world,
    we introduce a new audio deepfake dataset\footnote{\href{https://huggingface.co/datasets/mueller91/In-The-Wild}{https://huggingface.co/datasets/mueller91/In-The-Wild}}.
    We collect $17.2$ hours of high-quality audio deepfakes and $20.7$ hours of of authentic material from $58$ politicians and celebrities.
    \item We show that established models generally perform poorly on such real-world data. This discrepancy
between reported and actual generalization ability suggests that the detection of audio fakes is a far more difficult challenge than previously thought.

\end{itemize}

\section{Related Work}\label{sec:related_work}


\subsection{Model Architectures}

There is a significant body of work on audio spoof detection, driven largely by the ASVspoof challenges and datasets~\cite{todisco2019asvspoof, nautsch2021ASVspoof}.
In this section, we briefly present the architectures and models used in our evaluation in~\Cref{sec:results}.

\textbf{LSTM-based models}. Recurrent architectures are a natural choice in the area of language processing, with numerous related work utilizing such models~\cite{gomez-alanis2019Gated, chintha2020Recurrent, zhang2021initial, tambe2021deep}.
As a baseline for evaluating this approach, we implement a simple \lstm/ model:
it consists of three LSTM layers followed by a single linear layer.
The output is averaged over the time dimension to obtain a single embedding vector.

\textbf{LCNN}.
Another common architecture for audio spoof detection are LCNN-based learning models such as \lcnn/, \lcnna/, and \lcnnl/ \cite{wang2021Comparative,lavrentyeva_audio_2017,lavrentyeva_stc_2019}.
LCNNs combine convolutional layers with Max-Feature-Map activations to create `light' convolutional neural networks. \lcnna/ has an added single-head-attention pooling layer, while \lcnnl/ uses a Bi-LSTM layer and a skip connection.

\textbf{MesoNet}.
\meso/ is based on the Meso-4 \cite{afchar2018MesoNet} architecture, which was originally used for detecting facial video deepfakes.
It uses $4$ convolutional layers in addition to Batch Normalization, Max Pooling, and a fully connected classifier.

\textbf{MesoInception}.
Based on the facial deepfake detector Meso-Inception-4 \cite{afchar2018MesoNet}, \mi/ extends the Meso-4 architecture with Inception blocks \cite{inception}.

\textbf{ResNet18}. Residual Networks were first used for audio deepfake detection by~\cite{alzantot2019Deep}, and continue to be employed~\cite{ zhang2021one,monteiro2020generalized}. 
This architecture, first introduced in the computer vision domain~\cite{he2016deep}, uses convolutional layers and shortcut connections, which avoids the vanishing gradient problem and allows to design especially deep networks ($18$ layers for \resnet/).

\textbf{Transformer}. The Transformer architecture has also found its way into the field of audio spoof detection~\cite{zhang2021fake}. We use four self-attention layers with $256$ hidden dimensions and skip-connections, and encode time with positional encodings~\cite{vaswani2017attention}.

\textbf{CRNNSpoof}. This end-to-end architecture combines 1D convolutions with recurrent layers to learn features directly from raw audio samples~\cite{chintha2020Recurrent}.

\textbf{RawNet2}~\cite{tak2021EndtoEnd} is another end-to-end model. It employs Sinc-Layers~\cite{ravanelli2018speaker}, which correspond to rectangular band-pass filters, to extract information directly from raw waveforms.

\textbf{RawPC} is an end-to-end model which also uses Sinc-layers to operate directly on raw wavforms.
The architecture is found via differentiable architecture search~\cite{ge2021raw}.

\textbf{RawGAT-ST}, a spectro-temporal graph attention network (GAT), trained in an end-to-end fashion.
It introduces spectral and temporal sub-graphs and a graph pooling strategy, and reports state-of-the-art spoof detection capabilities~\cite{tak2021end}, which we can verify experimentally, c.f.~\Cref{tab:results}.

\section{Datasets}
\label{sec:dataset}

To train and evaluate our models, we use the ASVspoof 2019 dataset~\cite{todisco2019asvspoof}, in particular its \emph{Logical Access} (LA) part.
It consists of audio files that are either real (i.e., authentic recordings of human speech) or fake (i.e., synthesized or faked audio).
The spoofed audio files are from 19 different TTS synthesis algorithms.
From a spoofing detection point of view, ASVspoof considers synthetic utterances as a threat to the authenticity of the human voice, and therefore labels them as `attacks'.
In total, there are 19 different attackers in the ASVspoof 2019 dataset, labeled A1 - A19.
For each attacker, there are 4914 synthetic audio recordings and 7355 real samples.
This dataset is arguably the best known audio deefake dataset used by almost all related work.

\begin{figure}
    \centering
    \includegraphics[width=0.45\textwidth]{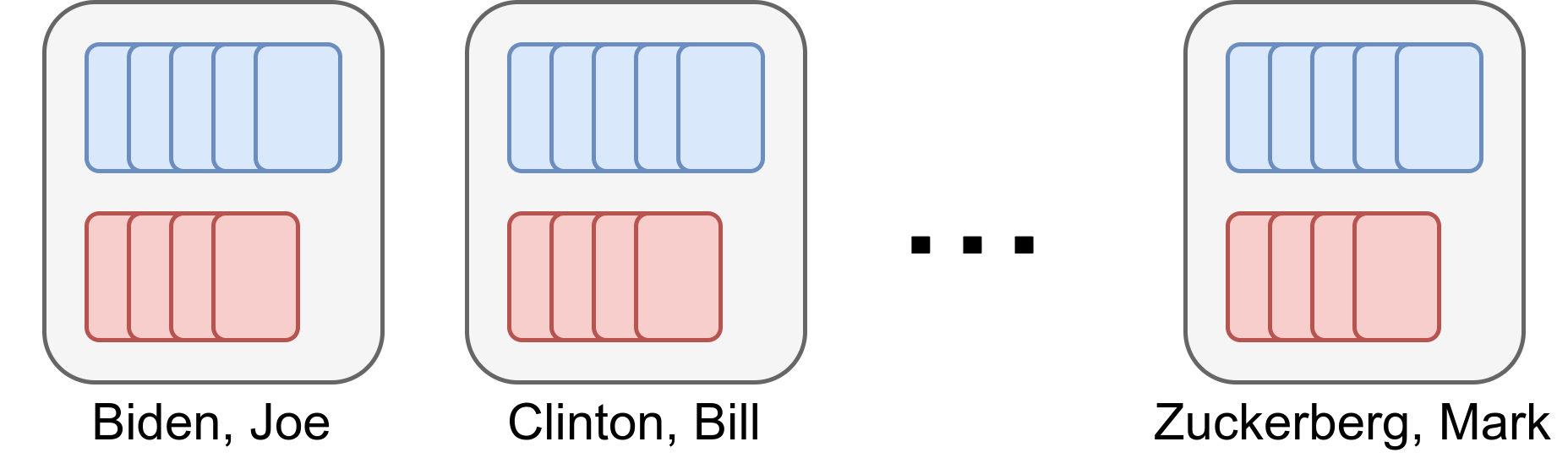}
    \caption{Schematics of our collected dataset. For $n=58$ celebrities and politicians, we collected both bona-fide and spoofed audio (represented by blue and red boxes per speaker).
    In total, we collected $20.8$ hours of bona-fide and $17.2$ hours of spoofed audio.
    On average, there are $23$ minutes of bona-fide and $18$ minutes of spoofed audio per speaker.
    }
    \label{fig:dataset_schema}
\end{figure}

In order to evaluate our models on realistic unseen data in-the-wild, we additionally create and publish a new audio deefake dataset, c.f.~\Cref{fig:dataset_schema}.
It consists of $37.9$ hours of audio clips that are either fake ($17.2$ hours) or real ($20.7$ hours).
We feature English-speaking celebrities and politicians, both from present and past\footnote{available at \href{https://huggingface.co/datasets/mueller91/In-The-Wild}{https://huggingface.co/datasets/mueller91/In-The-Wild}}.
The fake clips are created by segmenting $219$ of publicly available video and audio files that explicitly advertise audio deepfakes.
Since the speakers talk absurdly and out-of-character (`Donald Trump reads Star Wars'), it is easy to verify that the audio files are really spoofed.
We then manually collect corresponding genuine instances from the same speakers using publicly available material such as podcasts, speeches, etc.
We take care to include clips where the type of speaker, style, emotions, etc. are similar to the fake (e.g., for a fake speech by Barack Obama, we include an authentic speech and try to find similar values for background noise, emotions, duration, etc.).
The clips have an average length of $4.3$ seconds and are converted to `wav' after downloading.
All recordings were 
downsampled to 16 kHz (the highest common frequency in the original recordings).
Clips were collected from publicly available sources such as social networks and popular video sharing platforms.
This dataset is intended as evaluation data: it allows evaluation of a model's cross-database capabilities on a realistic use case.

\section{Experimental Setup}

\subsection{Training and Evaluation}

\begin{table*}
\centering
\footnotesize
\begin{tabular}{llrrrr}
\toprule
                    &              &      &  \multicolumn{2}{c}{ASVspoof19 eval}  & In-the-Wild Data\\
         Model Name & Feature Type & Input Length &   EER\% &     t-DCF & EER\% \\
\midrule
          \lcnn/ &      cqtspec &             Full &         \textbf{6.354±0.39} & 0.174±0.03 &        \textbf{65.559±11.14} \\
          \lcnn/ &      cqtspec &              4s &        25.534±0.10 & 0.512±0.00 &         70.015±4.74 \\
          \lcnn/ &      logspec &             Full &         7.537±0.42 & \textbf{0.141±0.02} &         72.515±2.15 \\
          \lcnn/ &      logspec &              4s &        22.271±2.36 & 0.377±0.01 &         91.110±2.17 \\
          \lcnn/ &      melspec &             Full &        15.093±2.73 & 0.428±0.05 &         70.311±2.15 \\
          \lcnn/ &      melspec &              4s &        30.258±3.38 & 0.503±0.04 &         81.942±3.50 \\
          \midrule
\lcnna/ &      cqtspec &             Full &         \textbf{6.762±0.27} & \textbf{0.178±0.01} &         \textbf{66.684±1.08} \\
\lcnna/ &      cqtspec &              4s &        23.228±3.98 & 0.468±0.06 &         75.317±8.25 \\
\lcnna/ &      logspec &             Full &         7.888±0.57 & 0.180±0.05 &         77.122±4.91 \\
\lcnna/ &      logspec &              4s &        14.958±2.37 & 0.354±0.03 &         80.651±6.14 \\
\lcnna/ &      melspec &             Full &        13.487±5.59 & 0.374±0.14 &         70.986±9.73 \\
\lcnna/ &      melspec &              4s &        19.534±2.57 & 0.449±0.02 &         85.118±1.01 \\
\midrule
     \lcnnl/ &      cqtspec &             Full &         \textbf{6.228±0.50} & \textbf{0.113±0.01} &         \textbf{61.500±1.37} \\
     \lcnnl/ &      cqtspec &              4s &        20.857±0.14 & 0.478±0.01 &         72.251±2.97 \\
     \lcnnl/ &      logspec &             Full &         9.936±1.74 & 0.158±0.01 &         79.109±0.84 \\
     \lcnnl/ &      logspec &              4s &        13.018±3.08 & 0.330±0.05 &        79.706±15.80 \\
     \lcnnl/ &      melspec &             Full &         9.260±1.33 & 0.240±0.04 &        62.304±0.17 \\
     \lcnnl/ &      melspec &              4s &        27.948±4.64 & 0.483±0.03 &         82.857±3.49 \\
     \midrule
               \lstm/ &      cqtspec &             Full &         \textbf{7.162±0.27} & \textbf{0.127±0.00} &        \textbf{53.711±11.68} \\
               \lstm/ &      cqtspec &              4s &        14.409±2.19 & 0.382±0.05 &         55.880±0.88 \\
               \lstm/ &      logspec &             Full &        10.314±0.81 & 0.160±0.00 &         73.111±2.52 \\
               \lstm/ &      logspec &              4s &        23.232±0.32 & 0.512±0.00 &         78.071±0.49 \\
               \lstm/ &      melspec &             Full &        16.216±2.92 & 0.358±0.00 &         65.957±7.70 \\
               \lstm/ &      melspec &              4s &        37.463±0.46 & 0.553±0.01 &         64.297±2.23 \\
               \midrule
     \mi/ &      cqtspec &             Full &        11.353±1.00 & 0.326±0.03 &        50.007±14.69 \\
     \mi/ &      cqtspec &              4s &        21.973±4.96 & 0.453±0.09 &        68.192±12.47 \\
     \mi/ &      logspec &             Full &        \textbf{10.019±0.18} & \textbf{0.238±0.02} &         \textbf{37.414±9.16} \\
     \mi/ &      logspec &              4s &        16.377±3.72 & 0.375±0.09 &         72.753±6.62 \\
     \mi/ &      melspec &             Full &        14.058±5.67 & 0.331±0.11 &        61.996±12.65 \\
     \mi/ &      melspec &              4s &        21.484±3.51 & 0.408±0.03 &        51.980±15.32 \\
     \midrule
           \meso/ &      cqtspec &             Full &         \textbf{7.422±1.61} & 0.219±0.07 &        54.544±11.50 \\
           \meso/ &      cqtspec &              4s &        20.395±2.03 & 0.426±0.06 &         65.928±2.57 \\
           \meso/ &      logspec &             Full &         8.369±1.06 & \textbf{0.170±0.05} &         \textbf{46.939±5.81} \\
           \meso/ &      logspec &              4s &        11.124±0.79 & 0.263±0.03 &        80.707±12.03 \\
           \meso/ &      melspec &             Full &        11.305±1.80 & 0.321±0.06 &        58.405±11.28 \\
           \meso/ &      melspec &              4s &        21.761±0.26 & 0.467±0.00 &        64.415±15.68 \\
           \midrule
    \resnet/ &      cqtspec &             Full &           \textbf{6.552±0.49} & \textbf{0.140±0.01} &                          \textbf{49.759±0.17} \\
   \resnet/ &      cqtspec &              4s &          18.378±1.76 & 0.432±0.07 &                          61.827±7.46 \\
   \resnet/ &      logspec &             Full &           7.386±0.42 & \textbf{0.139±0.02} &                          80.212±0.23 \\
   \resnet/ &      logspec &              4s &          15.521±1.83 & 0.387±0.02 &                          88.729±2.88 \\
   \resnet/ &      melspec &             Full &          21.658±2.56 & 0.551±0.04 &                          77.614±1.47 \\
   \resnet/ &      melspec &              4s &          28.178±0.33 & 0.489±0.01 &                          83.006±7.17 \\
    \midrule
    \transformer/ &      cqtspec &             Full &           \textbf{7.498±0.34} & \textbf{0.129±0.01} &                          \textbf{43.775±2.85} \\
    \transformer/ &      cqtspec &              4s &          11.256±0.07 & 0.329±0.00 &                          48.208±1.49 \\
    \transformer/ &      logspec &             Full &           9.949±1.77 & 0.210±0.06 &                          64.789±0.88 \\
    \transformer/ &      logspec &              4s &          13.935±1.70 & 0.320±0.03 &                          44.406±2.17 \\
    \transformer/ &      melspec &             Full &          20.813±6.44 & 0.394±0.10 &                          73.307±2.81 \\
    \transformer/ &      melspec &              4s &          26.495±1.76 & 0.495±0.00 &                          68.407±5.53 \\
           \midrule
           \midrule
           \crnn/ &          raw &             Full &        \textbf{15.658±0.35} & \textbf{0.312±0.01} &         44.500±8.13 \\
           \crnn/ &          raw &              4s &        19.640±1.62 & 0.360±0.04 &         \textbf{41.710±4.86} \\
           \midrule
            \rn/ &          raw &             Full &         \textbf{3.154±0.87} & \textbf{0.078±0.02} &         37.819±2.23  \\
            \rn/ &          raw &              4s &         4.351±0.29 & 0.132±0.01 &         \textbf{33.943±2.59} \\
           \midrule
            \RawPC/ &         raw &           Full & \textbf{3.092±0.36}  & \textbf{0.071±0.00} & \textbf{45.715±12.20}  \\
            \RawPC/ &         raw &           4s &  {3.067±0.91}   &   {0.097±0.03}      & {52.884±6.08} \\
           \midrule
            \rawgatst/ &         raw &           Full & \textbf{1.229±0.43}       & \textbf{0.036±0.01} & \textbf{37.154±1.95} \\
            \rawgatst/ &          raw &              4s & {2.297±0.98}   &   {0.074±0.03} & {38.767±1.28} \\
\bottomrule
\end{tabular}

\vspace*{0.3cm}
\caption{Full results of evaluation on the ASVspoof 2019 LA `eval' data.
We compare different model architectures against different feature types and audio input lengths (4s, fixed-sized inputs vs. variable-length inputs).
Results are averaged over \runs/ independent trials with random initialization, and the standard deviation is reported. 
Best-performing configurations are highlighted in boldface. 
When evaluating the models on our proposed `in-the-wild' dataset, we see an increase in EER by 
up to 1000\% compared to ASVspoof 2019 (rightmost column).
}
\label{tab:results}
\end{table*}


\subsubsection{Hyper Parameters}
We train all of our models using a cross-entropy loss with a log-Softmax over the output logits.
We choose the Adam \cite{kingmaAdam} optimizer.
We initialize the learning rate at $0.0001$ and use a learning rate scheduler.
We train for $100$ epochs with early stopping using a patience of five epochs.

\subsubsection{Train and Evaluation Data Splits}
We train our models on the `train' and `dev' parts of the ASVSpoof 2019 Logical Access (LA) dataset part~\cite{todisco2019asvspoof}.
This is consistent with most related work and also with the evaluation procedure of the ASVspoof 2019 Challenge.
We test against two evaluation datasets. 
As \emph{in-domain} evaluation data, we use the `eval' split of ASVspoof 2019.
This split contains unseen attacks, i.e., attacks not seen during training.
However, the evaluation audios share certain properties with the training data~\cite{muller2021Speech}, so model generalization cannot be assessed using the `eval' split of ASVspoof 2019 alone.
This motivates the use of our proposed `in-the-wild' dataset, see Section~\ref{sec:dataset}, as unknown \emph{out-of-domain} evaluation data.

\subsubsection{Evaluation metrics}
We report both the equal-error rate (EER) and the tandem detection cost function (t-DCF) \cite{kinnunen_t-dcf_2018} on the ASVspoof 2019 `eval' data.
For consistency with the related work, we use the original implementation of the t-DCF as provided for the ASVspoof 2019 challenge~\cite{tdcf:online}.
For our proposed dataset, we report only the EER.
This is because t-DCF scores require the false alarm and miss costs, which are available only for ASVspoof.

\subsection{Feature Extraction}

Several architectures used in this work require pre-processing the audio data with a feature extractor (\lcnn/, \lcnna/, \lcnnl/, \lstm/, \meso/, \mi/, \resnet/, \transformer/).
We evaluate these architectures on constant-Q transform (\emph{cqtspec}~\cite{brown1991calculation}), log spectrogram (\emph{logspec}) and mel-scaled spectrogram (\emph{melspec}~\cite{stevens1937scale}) features (all of them 513-dimensional).
We use Python, \emph{librosa}~\cite{mcfee2015librosa} and \emph{scipy}~\cite{virtanen2020scipy}.
The rest of the models does not rely on pre-processed data, but uses raw audio waveforms as inputs.

\subsection{Audio Input Length}

Audio samples usually vary in length, which is also the case for the data in ASVspoof 2019 and our proposed `in-the-wild' dataset.
While some models can accommodate variable-length input (and thus also fixed-length input), many can not.
We extend these by introducing a global averaging layer, which adds such capability. 


In our evaluation of fixed-length input, we chose a length of four seconds, following~\cite{tak2021EndtoEnd}.
If an input sample is longer, a random four-second subset of the sample is used. 
If it is shorter, the sample is repeated.
To keep the evaluation fair, these shorter samples are also repeated during the full-length evaluation.
This ensures that full-length input is never shorter than truncated input, but always at least $4$s.



\section{Results}
\label{sec:results}



Table~\ref{tab:results} shows the results of our experiments, where we evaluate all models against all configurations of data preprocessing: we train \nummodels/ different models, using one of four different feature types, with two different ways of handling variable-length audio.
Each experiment is performed \runs/ times, using random initialization.
We report averaged EER and t-DCF, as well as standard deviation.
We observe that on ASVspoof, our implementations perform comparable to related work, with a margin
of approximately $2 - 4\%$ EER and $0.1$ t-DCF. 
This is likely because we do not fine-tune our models' hyper-parameters.

\subsection{Fixed vs. Variable Input Length}
We analyze the effects of truncating the input signal to a fixed length compared to using the full, unabridged audio.
For all models, performance decreases when the input is trimmed to $4s$.
Table~\ref{tab:trim-subsel} averages all results based on input length.
We see that average EER on ASVspoof drops from $19.89$\% to $9.85$\% when the full-length input is used.
These results show that a four-second clip is insufficient for the model to extract useful information compared to using the full audio file as input.
Therefore, we propose not to use fixed-length truncated inputs, but to provide the full audio file to the model.
This may seem obvious, but the numerous works that use fixed-length inputs~\cite{tak2021EndtoEnd, ge2021raw, tak2021end} suggest otherwise.

\begin{table}
\centering
\begin{tabular}{lrrr}
\toprule
     &  \multicolumn{2}{c}{ASVspoof19 eval} &     In-the-Wild Data \\
Input Length & EER \%       & t-DCF      &  EER \%  \\
\midrule
 Full &         9.85 &  0.22 &            60.10 \\
  4s &         18.89 &  0.39 &            67.25 \\
\bottomrule
\end{tabular}

\vspace*{0.3cm}
\caption{Model performance averaged by input preprocessing.
Fixed-length, 4s inputs perform significantly worse on the ASVspoof data and on the `in-the-wild' dataset than variable-length inputs.
This suggests that related work using fixed-length inputs may (unnecessarily) sacrifice performance.
}
\label{tab:trim-subsel}
\end{table}




\subsection{Effects of Feature Extraction Techniques}
We discuss the effects of different feature preprocessing techniques, c.f.~\ref{tab:results}:
The `raw' models outperform the feature-based models, obtaining up to $1.2\%$ EER on ASVspoof and $33.9\%$ EER on the `in-the-wild' dataset (\rawgatst/ and \rn/).
The spectrogram-based models perform slightly worse, achieving up to $6.3\%$ EER on ASVspoof and $37.4\%$ on the `in-the-wild' dataset (\lcnn/ and \meso/).
The superiority of the `raw' models is assumed to be due to finer feature-extraction resolution than the spectogram-based models~\cite{tak2021end}.
This has lead recent research to focus largely on such raw-feature, end-to-end models~\cite{ge2021raw, tak2021end}.

Concerning the spectogram-based models, we observe that \emph{melspec} features are always outperformed by either \emph{cqtspec} of \emph{logspec}.
Simply replacing \emph{melspec} with \emph{cqtspec} increases the average performance by 37\%, all other factors constant.


\subsection{Evaluation on `in-the-wild' data}
Especially interesting is the performance of the models on real-world deepfake data.
Table~\ref{tab:results} shows the performance of our models on the `in-the-wild' dataset.
We see that there is a large performance gap between the ASVSpoof 2019 evaluation data and our proposed `in-the-wild' dataset.
In general, the EER values of the models deteriorate by about 200 to 1000 percent. 
Often, the models do not perform better than random guessing.

To investigate this further, we train our best `in-the-wild' model from \Cref{tab:results}, \rn/ with $4s$ input length, on \emph{all} from ASVspoof 2019, i.e., the `train', `dev', and `eval' splits.
We then re-evaluate on the `in-the-wild' dataset to investigate whether adding more ASVspoof training data improves out-of-domain performance.
We achieve $33.1 \pm 0.2$~\%~EER, i.e., no improvement over training with only the `train' and `dev' data.

The inclusion of the `eval' split does not seem to add much information that could be used for real-world generalization. 
This is plausible in that all splits of ASVspoof are fundamentally based on the same dataset, VCTK, although the synthesis algorithms and speakers differ between splits~\cite{todisco2019asvspoof}.

\section{Conclusion}
In this paper, we systematically evaluate audio spoof detection models from related work according to common standards.
In addition, we present a new audio deefake dataset of `in-the-wild' audio spoofs that we use to evaluate the generalization capabilities of related work in a real-world scenario.

We find that regardless of the model architecture, some preprocessing steps are more successful than others.
It turns out that the use of \emph{cqtspec} or \emph{logspec} features consistently outperforms the use of \emph{melspec} features in our comprehensive analysis.
Furthermore, we find that for most models, four seconds of input audio does not saturate performance compared to longer examples.
Therefore, we argue that one should consider using \emph{cqtspec} features and unabridged input audio when designing audio deepfake detection architectures.

Most importantly, however, we find that the `in-the-wild' generalization capabilities of many models may have been overestimated.
We demonstrate this by collecting our own audio deepfake dataset and evaluating \nummodels/ different model architectures on it. 
Performance drops sharply, and some models degenerate to random guessing.
It may be possible that the community has tailored its detection models too closely to the prevailing benchmark, ASVSpoof, and that deepfakes are much harder to detect outside the lab than previously thought.

\bibliographystyle{IEEEtran}
\bibliography{paper}

\end{document}